\documentclass[%
 reprint,
 amsmath,amssymb,
 aps,
]{revtex4-2}

\usepackage{graphicx}
\usepackage{dcolumn}
\usepackage{bm}
\usepackage{siunitx}
\usepackage{amsmath}
\usepackage[english]{babel}

\usepackage{svg}

\begin{document}

\preprint{APS/123-QED}
\twocolumngrid

\title{Gigahertz-Frequency Acousto-Optic Phase Modulation of Visible Light in a CMOS-Fabricated Photonic Circuit}

\author{Jacob M. Freedman$^1$}
 \author{Matthew J. Storey$^2$}
 \author{Daniel Dominguez$^2$}
 \author{Andrew Leenheer$^2$} 
 \author{Sebastian Magri$^1$}
 \author{Nils T. Otterstrom\email{ntotter@sandia.gov}$^{2}$}\email{ntotter@sandia.gov}
 \author{Matt Eichenfield$^{1,2}$}\email{eichenfield@arizona.edu}

\affiliation{\phantom{$^1$ James C. Wyant College of Optical Sciences, University of Arizona, Tucson, Arizona, USA} \\ $^1$ Wyant College of Optical Sciences, University of Arizona, Tucson, Arizona, USA \\ $^2$ Microsystems Engineering, Science, and Applications, Sandia National Laboratories, Albuquerque, New Mexico, USA
}

\begin{abstract}
Acousto-optic devices are broadly utilized in optical signal processing to modulate light's amplitude, phase, frequency, and propagation direction. The core functions of acousto-optic technologies include high-extinction pulse carving, single-sideband frequency shifting, beam deflection, and tunable filtering. Recent advances in the chip-scale implementation of these devices have allowed for higher modulation frequencies, miniaturization of device footprint, and reduction of microwave power consumption. This on-chip optical signal processing allows for a single function to be arrayed at scale over many simultaneously controllable channels derived from a single laser input.  Yet, these demonstrations have primarily operated at telecom wavelengths or made use of materials that impede scalability such as lithium niobate, which suffers from poor cross-wafer uniformity, low power-handling capability, and CMOS-incompatibility. Here we present an efficient, visible-light, gigahertz-frequency acousto-optic modulator fabricated on a $\SI{200}{\milli\metre}$ wafer in a volume CMOS foundry. Our device combines a piezoelectric transducer and a photonic waveguide within a single microstructure that confines both a propagating optical mode and an electrically excitable breathing-mode mechanical resonance. By tuning the device's geometry to optimize the optomechanical interaction, we achieve modulation depths exceeding $\SI{2}{\radian}$ with $\SI{15}{\milli\watt}$ applied microwave power at $\SI{2.31}{\giga\hertz}$ in a $\SI{2}{\milli\metre}$ long device. This corresponds to a modulation figure of merit of $V_{\pi}\cdot L = \SI{0.26}{\volt\centi\meter}$ in a visible-light, integrated acousto-optics platform that can be straightforwardly extended to a wide range of optical wavelengths and modulation frequencies. For the important class of gigahertz-frequency modulators that can handle hundreds of milliwatts of visible-light optical power, which are critical for scalable quantum control systems, this represents a 15x decrease in $V_{\pi}$ and a 100x decrease in required microwave power compared to the commercial state-of-the-art and existing work in the literature.
\end{abstract}

\maketitle

\section{\label{sec:level1}INTRODUCTION}

\begin{figure*}
    \includegraphics[width=18cm]{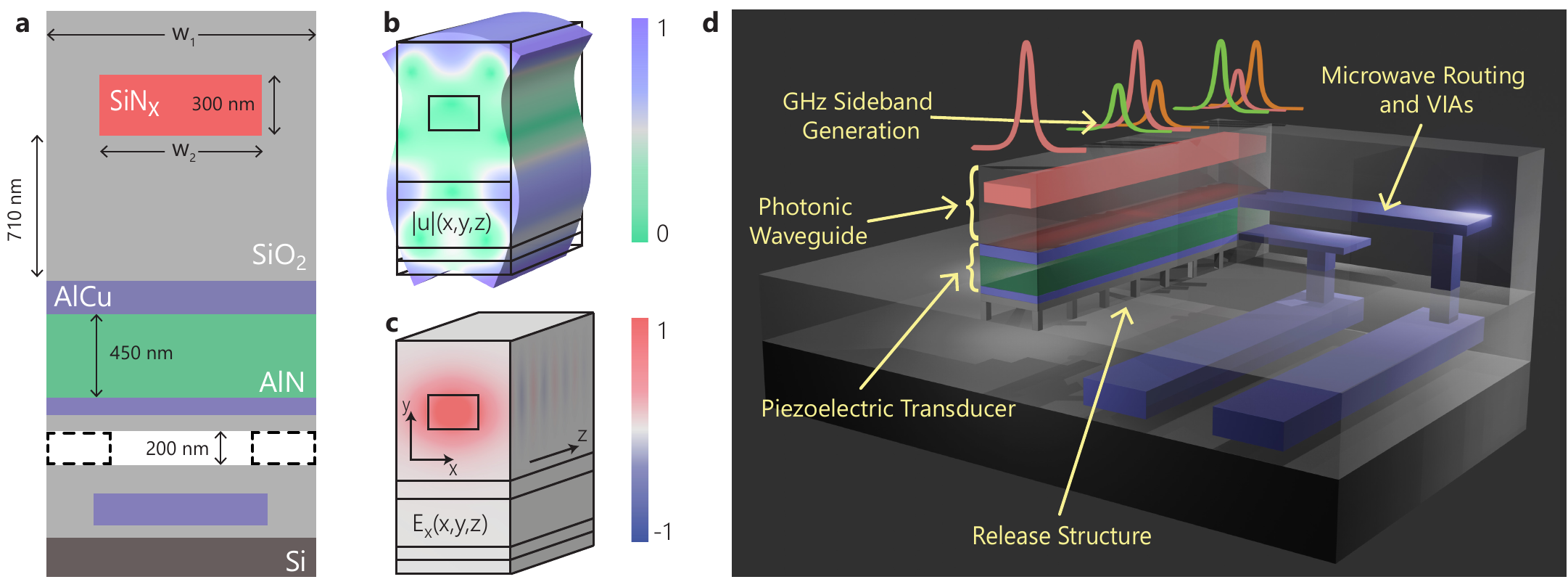}
    \caption{\label{fig:wide}{\bf Design of the CMOS-fabricated, resonantly enhanced acousto-optic modulator.} {\bf a} Cross-section of the modulator showing the material stack and layer thicknesses. Resonant mechanical breathing modes of the structure are piezoelectrically excited by applying a microwave voltage across the electrodes that sandwich the AlN film. These mechanical deformations modify the effective refractive index of the device's optical mode and consequently modulate its phase. The device is released from the substrate except for periodically placed supports whose locations are indicated by dashed boxes. {\bf b} Simulated profile of a $\SI{2.27}{\giga\hertz}$ mechanical mode's normalized displacement magnitude. {\bf c} Simulated normalized $x$-component of electric field for the waveguide's fundamental tranverse electric mode at $\SI{730}{\nano\metre}$, to which the mechanical mode is strongly optomechanically coupled. The optical mode is designed to be far enough from the metal electrode to avoid absorption loss, yet close enough to significantly overlap with electromechanically excited strain-fields. {\bf d} Three-dimensional graphical representation of the device. An etch and release-layer remove material to the sides of and below the modulator, which is supported by periodically placed nanopillars. Microwaves are coupled to the electrodes of the device's piezoelectric transducer through a routing layer and vertical interconnect accesses.
    }
\end{figure*}

Bulk acousto-optic technology has enabled high-extinction pulse carving, single-sideband frequency shifting, beam deflection, and tunable filtering \cite{savage_acousto-optic_2010}. However, the devices that provide these critical functions typically consume high power and operate at low frequencies. Recently, on-chip acousto-optics platforms have yielded fast optical switches \cite{dong_high-speed_2022}, microwave-to-optical transducers \cite{shao_microwave--optical_2019, blesin_bidirectional_2024, jiang_efficient_2020}, isolators \cite{tian_magnetic-free_2021, sohn_electrically_2021, peterson_synthetic_2018}, beam deflectors \cite{li_frequencyangular_2023}, and narrowband filters \cite{gertler_narrowband_2022}. These devices offer increasingly compact and energy-efficient versions of their traditionally large, high-power-consumption bulk analogues and facilitate higher frequency operation in the gigahertz range \cite{shao_microwave--optical_2019, li_electromechanical_2019, savage_acousto-optic_2010}. Also, because non-vanishing photoelastic coefficients exist regardless of a material's crystal symmetries, acousto-optic devices can be implemented in a significantly broader class of materials than their electro-optic counterparts \cite{mueller_theory_1935}. This flexibility allows for the use of very large scale integration-compatible materials such as silicon nitride (SiN$_x$) that have desirable properties including wide transparency window \cite{xiang_silicon_2022}, low propagation loss \cite{parsons_low_1991}, high power-handling capability \cite{blumenthal_silicon_2018}, and compatibility with complementary metal oxide semiconductor (CMOS) fabrication processes \cite{moss_new_2013}.

Demonstrations of high-frequency, integrated electro-optic \cite{hu_-chip_2021, renaud_sub-1_2023, kodigala_high-performance_2024, yu_single-chip_2018} and acousto-optic \cite{erdil_wideband_2024, kittlaus_electrically_2021, tadesse_sub-optical_2014, zhang_integrated-waveguide-based_2024, shao_integrated_2020} frequency modulators with excellent performance have been implemented in a variety of material platforms. Yet to date, these operate with telecom-band light or else rely on materials that are both incompatible with CMOS foundry processes and incapable of supporting high optical power \cite{zhu_integrated_2021}. This limits their applicability in optical systems operating with high power at visible wavelengths such as those for underwater communication \cite{notaros_liquid-crystal-based_2023}, visible-light laser ranging (LiDAR) \cite{jantzi_enhanced_2018}, and quantum control \cite{bluvstein_quantum_2022}. For the latter, leveraging a SiN$_x$ platform facilitates integration with previously demonstrated functionality such as stable splitting and routing \cite{mehta_integrated_2020, kwon_multi-site_2024}, chip-to-qubit free-space coupling with focusing diffraction gratings \cite{mehta_precise_2017}, polarization rotators \cite{hattori_integrated_2024}, and active phase shifters \cite{dong_high-speed_2022}. The addition of on-chip optical frequency control to this growing library of components would serve as a crucial resource for generating individually addressable, channelized Raman gates and cooling beams, which require visible-light operation and gigahertz frequency shifts \cite{bruzewicz_trapped-ion_2019, lim_design_2025}. 

In this work, we present the design and experimental realization of a CMOS-fabricated, resonant acousto-optic phase modulator implemented in a piezo-optomechanical aluminum nitride-SiN$_x$ platform. We obtain a strong optomechanical interaction by confining a propagating optical mode and a breathing-mode mechanical resonance to a single, wavelength-scale structure. A piezoelectric actuator embedded within the structure, combined with this optomechanical coupling, enables efficient electromechanically driven acousto-optic phase modulation. Our device harnesses this phase modulation to generate gigahertz-frequency sidebands on a $\SI{730}{\nano\metre}$ laser input. At a $\SI{2.31}{\giga\hertz}$ modulation frequency, we achieve a modulation depth of $\SI{2.1}{\radian}$ with a $\SI{2}{\milli\metre}$ long device and $\SI{15}{\milli\watt}$ of applied microwave power. State-of-the-art electro-optic phase modulators capable of operating with high optical power at similar wavelengths typically utilize bulk lithium niobate (LN) and require watt-level microwave power to produce modulation depths of order $\SI{1}{\radian}$, corresponding to $V_{\pi} \approx \SI{20}{\volt}$ \cite{qubig_resonant_nodate, thorlabs_free_space}. These devices sacrifice efficiency for power handling because the photorefractive effect limits LN's maximum supportable optical power density, requiring large modal cross-sections to accommodate high power and necessitating a large separation between driving electrodes \cite{savchenkov_enhancement_2006}. Doping LN with magnesium oxide can mitigate photorefractive damage, but even with this, the maximum power density at visible wavelengths has been reported at $\SI{2}{\micro\watt/\micro\metre^2}$ \cite{thorlabs_free_space}. Extrapolating this value to the cross-section of a single mode waveguide in thin-film LN at visible wavelengths implies power handling well below a milliwatt. So, while waveguide-based, visible-light LN modulators achieve significantly lower $V_{\pi}$ than their bulk analogues, the tight confinement of light in this geometry likely limits optical powers well below milliwatt levels \cite{desiatov_ultra-low-loss_2019}. Our device utilizes SiN$_x$ waveguides with similar dimensions to those that have operated with hundreds of milliwatts of optical power at $\SI{729}{\nano\metre}$ \cite{mehta_integrated_2020} and $\SI{780}{\nano\metre}$ \cite{donvalkar_broadband_2017}, making it, to the best of our knowledge, the highest performing resonant phase modulator with high power handling at visible wavelengths. 

The key advantage of on-chip optical signal processing lies in its scalability, allowing many channels derived from a single, high-power laser to be simultaneously controlled. If each channel must deliver a critical modulated power to a target, then the photonics platform must have a power threshold that is some multiplicative factor of the application-specific critical power. Trapped ion and neutral atom quantum computers will require thousands or even millions of channels with independent frequency control for each qubit \cite{park_technologies_2024, bruzewicz_trapped-ion_2019}. Gate operations typically require tens of milliwatts of laser power incident on the qubit site \cite{moses_race-track_2023}, meaning integrated photonic control chips will need to handle watt-level power. Our device is uniquely capable of satisfying the scaling requirements for such atom and ion control systems, as it employs SiN$_x$ waveguides known for their power-handling at visible wavelengths, is fabricated on a $\SI{200}{\milli\metre}$ wafer in a CMOS foundry, has a 15x decrease in $V_{\pi}$ and a 100x reduction in required microwave power compared to the state-of-the-art, and is built on a platform in which many functions such as high-speed switches \cite{dong_high-speed_2022} and tunable narrowband filters \cite{stanfield_cmos-compatible_2019} already exist.  It further lays the foundation for a visible-light, acousto-optics platform capable of supporting a variety of components such as isolators and single-sideband frequency modulators. 

\begin{figure*}
    \includegraphics[width=17cm]{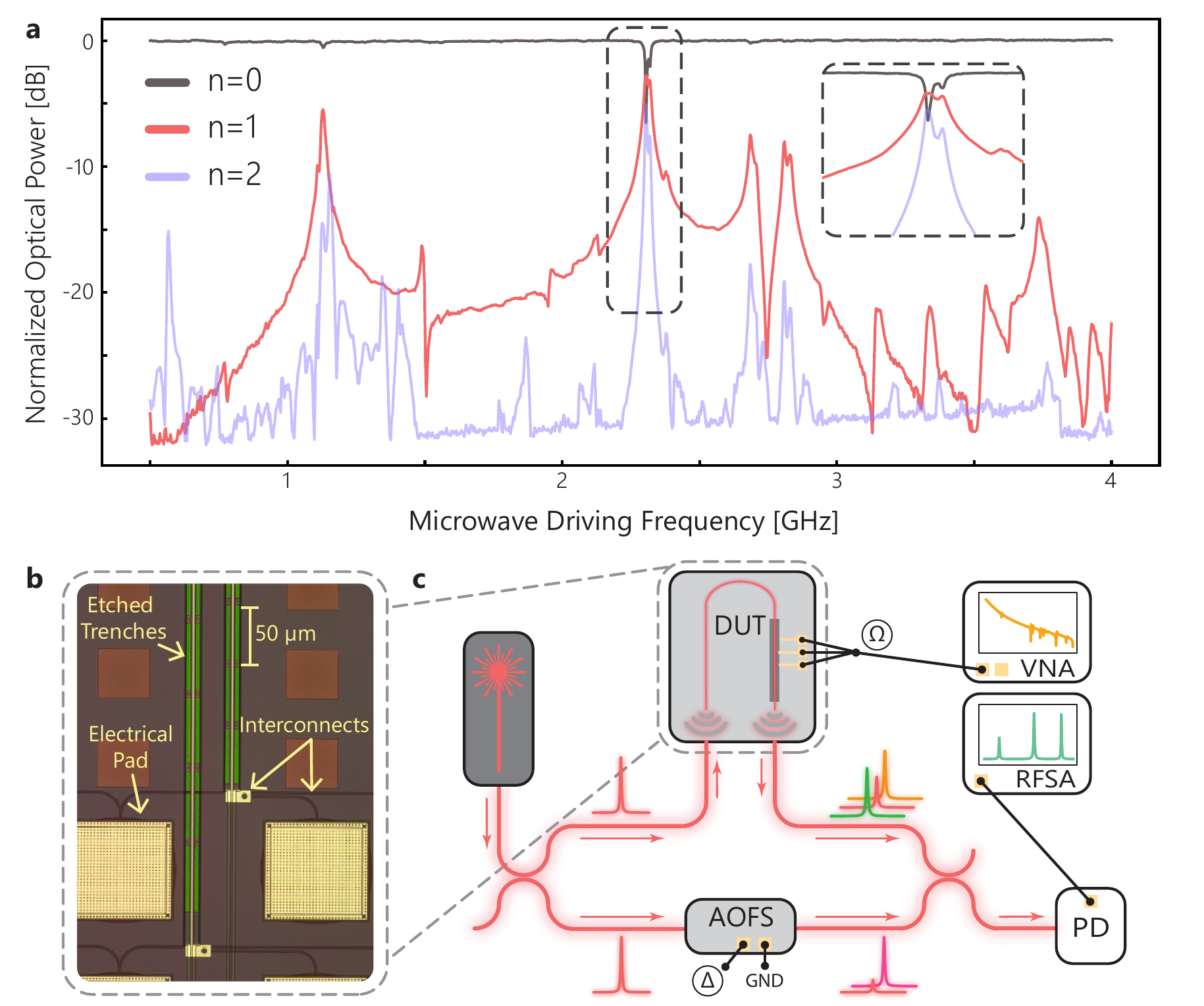}
    \caption{\label{fig:wide}{\bf Observation of on-chip, gigahertz-frequency sideband generation at visible wavelengths.} {\bf a} Plot of normalized optical power at the input laser frequency ($n=0$), first order sideband ($n=1$), and second order sideband ($n=2$) as a function of the microwave drive frequency $\Omega / 2\pi$. Resonantly enhanced phase-modulation is most prominently observed at $\SI{1.13}{\giga\hertz}$, $\SI{2.31}{\giga\hertz}$, $\SI{2.68}{\giga\hertz}$, and $\SI{2.80}{\giga\hertz}$. Inset shows the $\SI{2.31}{\giga\hertz}$ resonance in detail where the optical power of the first and second order sidebands each eclipse that of the remaining carrier. {\bf b} Micrograph of a representative device. Two separate phase modulators run from bottom to top, and the green channels indicate where an etch has structurally isolated the device from the die. The two large squares at the bottom are conductive pads through which microwaves are coupled to the chip, and routing lines can be seen emanating from these that deliver the microwaves to the devices' piezoelectric actuators.  {\bf c} Experimental diagram of the setup for characterizing phase modulation. Laser light is split into two arms of an off-chip fiber interferometer. Light from the upper arm is grating-coupled to and from the DUT, which receives microwave power at a frequency $\Omega / 2\pi$ from a VNA that also records the microwave power reflected off the device. Meanwhile, light from the lower arm is frequency-shifted by an amount $\Delta / 2\pi$. The arms are then interfered with a directional coupler to generate intensity modulation at angular frequencies $n\Omega \pm \Delta$, whose strengths are related to the optical power in the $n$th sideband produced by the phase modulation process. Finally, a fast photodetector converts the intensity modulation to an electric signal which is read out by a radio frequency spectrum analyzer.}
\end{figure*}

\section{\label{sec:level1}DEVICE PLATFORM AND WORKING PRINCIPLE}

Our acousto-optic modulator comprises a photonic waveguide situated directly above a piezoelectric actuator. A SiN$_x$ core enclosed by SiO$_2$ cladding forms the waveguide, and the actuator leverages the dominant $d_{33}$ and $d_{31}$ piezoelectric coefficients of an aluminum nitride film by sandwiching it between two electrode layers (Fig. 1a) \cite{lueng_piezoelectric_2000}. We utilize etched trenches and a release layer to physically isolate the structure from the wafer substrate and surrounding photonic and piezoelectric layers (Fig. 1d). This inhibits the principal mechanism by which mechanical energy can escape the device and consequently allows sharp mechanical resonances to exist.

We obtain resonantly enhanced optomechanical coupling by confining a breathing-mode mechanical resonance and a propagating optical mode to the resulting wavelength-scale structure. The coupling results from the photoelastic effect, which describes strain-induced changes to a material's refractive index, and the moving boundary effect, which accounts for alterations to a waveguide's cross-section by mechanical displacements at the core-cladding interface \cite{johnson_perturbation_2002, wolff_stimulated_2015}. These mechanically-driven effects modify the optical mode's effective refractive index by an amount $\Delta n_{\text{eff}}$. The optical mode then accumulates a phase shift of $\Delta\phi = \omega \Delta n_{\text{eff}} L / c$ after propagating through a length $L$ over which this index-modification is present, where $\omega$ is the angular frequency of the optical mode and $c$ is the speed of light in vacuum.

We electrically excite resonantly enhanced phase shifts by applying a sinusoidal microwave signal to a device's piezoelectric transducer at a mechanical resonance frequency, $\Omega$. The index modification then becomes $\Delta n_{\text{eff}}(t) = \Delta n_0 \sin (\Omega t)$, which produces modulation of the form
    \begin{align}
        E(t) = E_0\exp{\left\{i\left[-\omega t + \alpha\sin\left(\Omega t\right)\right]\right\}}, \label{eq:phasemod}
    \end{align}
where $\alpha = \omega\Delta n_0 L / c$ is the modulation depth resulting from a sinusoidally varying index modification with amplitude $\Delta n_0$. A phase mismatch between the optical mode and mechanically induced index perturbation constrains the above expression for $\alpha$ to the regime in which the optical traversal time of the modulator is much shorter than a period of the mechanical mode's oscillation (Supplement I). The device presented in this work has a length of $\SI{2}{\milli\metre}$ that falls within this regime.

Device performance, measured in modulation depth imparted per volt of applied microwave signal, ultimately depends on the entire piezo-optomehcanical transduction chain. First, microwave power is transferred from driving electronics to the on-chip device, where effective voltage drop across the device depends on its electrical impedance matching with the microwave transmission line. Second, the piezoelectric transducer converts some fraction of the delivered microwave power to mechanical energy, the efficiency of which depends on the electromechanical coupling coefficient, $k^2$ \cite{yan_near-ideal_2022}. Third, as discussed above, the optomechanical coupling between the two modes in question determines the change to the optical propagation constant per unit excited mechanical displacement. Finally, the quality factor, $Q$, of the excited mechanical mode increases the displacement and strain build-up in the structure and consequently determines the degree to which the modulation is resonantly enhanced.

Taking into account the above, we optimize our modulator's design by simulating the field profiles of the device's mechanical resonances and optical modes as a function of the cladding width, $w_1$, and SiN$_x$ width, $w_2$. From these field profiles we calculate the photoelastic and moving boundary coupling integrals given by
\begin{align}
    \Delta \beta_{\text{pe}} &= \kappa \iint\mathbf{E}^*\cdot\Delta\mathbf{\epsilon}\cdot\mathbf{E}\phantom{x}dA \label{eq:pe} \\[1.5em]
    \Delta \beta_{\text{mb}} &= \kappa \oint \mathbf{u}
    \cdot\hat{\mathbf{n}}\left[\Delta\epsilon_{12}|\mathbf{E}_{||}|^2-\Delta\epsilon_{12}^{-1}|\mathbf{D}_{\perp}|^2\right]dl \label{eq:mb}.
\end{align}
In equation \eqref{eq:pe}, the integral is evaluated over a cross-section of the device, $\mathbf{E}$ is the optical electric field profile, and $\Delta\epsilon$ is the photoelastically induced perturbation to the permittivity. In equation \eqref{eq:mb}, the integral is evaluated along the boundary between the SiN$_x$ and SiO$_2$, $\mathbf{u}$ is the mechanical displacement field, $\hat{\mathbf{n}}$ is the outward-facing normal vector to the boundary, $\mathbf{E}_{||}$ is the electric field parallel to the boundary, $\mathbf{D}_{\perp}$ is the electric displacement field perpendicular to the boundary, $\Delta\epsilon_{12}$ is the difference in permittivity between the SiO$_2$ and SiN$_x$, and $\Delta\epsilon^{-1}_{12}$ is the difference in reciprocal of the permittivity between the SiO$_2$ and SiN$_x$. In both equations $\kappa$ is a constant that includes normalization factors. The simulated field profiles of a strongly coupled mode-pairing are shown in Fig. 1b-c, and details of the coupling calculation, as well as simulations of $k^2$ are found in Supplement II. In principle, one should maximize the product of $k^2$, optomechanical coupling, microwave power transfer efficiency, and mechanical quality factor. However, $Q$ is difficult to accurately simulate and also critically affects the device's equivalent electrical impedance within a linewidth of a mechanical resonance \cite{larson_modified_2000}. In the absence of exact models to predict the mechanical $Q$, we design our devices to mitigate radiative loss to the substrate and surrounding material.

Phase modulators are often used to generate gigahertz-frequency sidebands, and achieving maximum conversion efficiency, defined here as the fraction of optical power converted to one of the first-order sidebands, is thus an important performance benchmark \cite{levine_dispersive_2022, lee_atomic_2003, delhaye_hybrid_2012}. By rewriting equation \eqref{eq:phasemod} with the Jacobi-Anger expansion as

    \begin{align}
        E(t) = E_0 \sum_{n=-\infty}^{\infty} J_n(\alpha)\exp{\left[-i\left(\omega - n\Omega\right)t \right]}, \label{eq:jacobi}
    \end{align}
    
one can see optical power in the first order sideband is maximized when $J_1(\alpha)$ itself is maximized, which occurs when $\alpha=\SI{1.84}{\radian}$ and yields a conversion efficiency of $\SI{33.9}{\percent}$. A large $Q$ is necessary to obtain maximum optical power in the first order sideband while maintaining low-voltage operation because the constraints on device length imposed by the phase-matching condition render it impossible to indefinitely elongate the device to achieve larger phase shifts. However some applications, particularly those for quantum control, require fast on/off time and do not necessarily favor high-$Q$ operation \cite{ivory_integrated_2021}. This is because increasing the quality factor prolongs the ring-up and decay time $\tau = 2Q/\Omega$ of the mechanical mode, which is the switching time of the modulator. Therefore, it is desirable to have strong enough optomechanical and electromechanical coupling so that a device can efficiently produce modulation depths near $\alpha = \SI{1.84}{\radian}$ without necessitating ultra-high quality factors that preclude fast switching times and limit the bandwidth of operation.

\begin{figure*}
    \includegraphics[width=17cm]{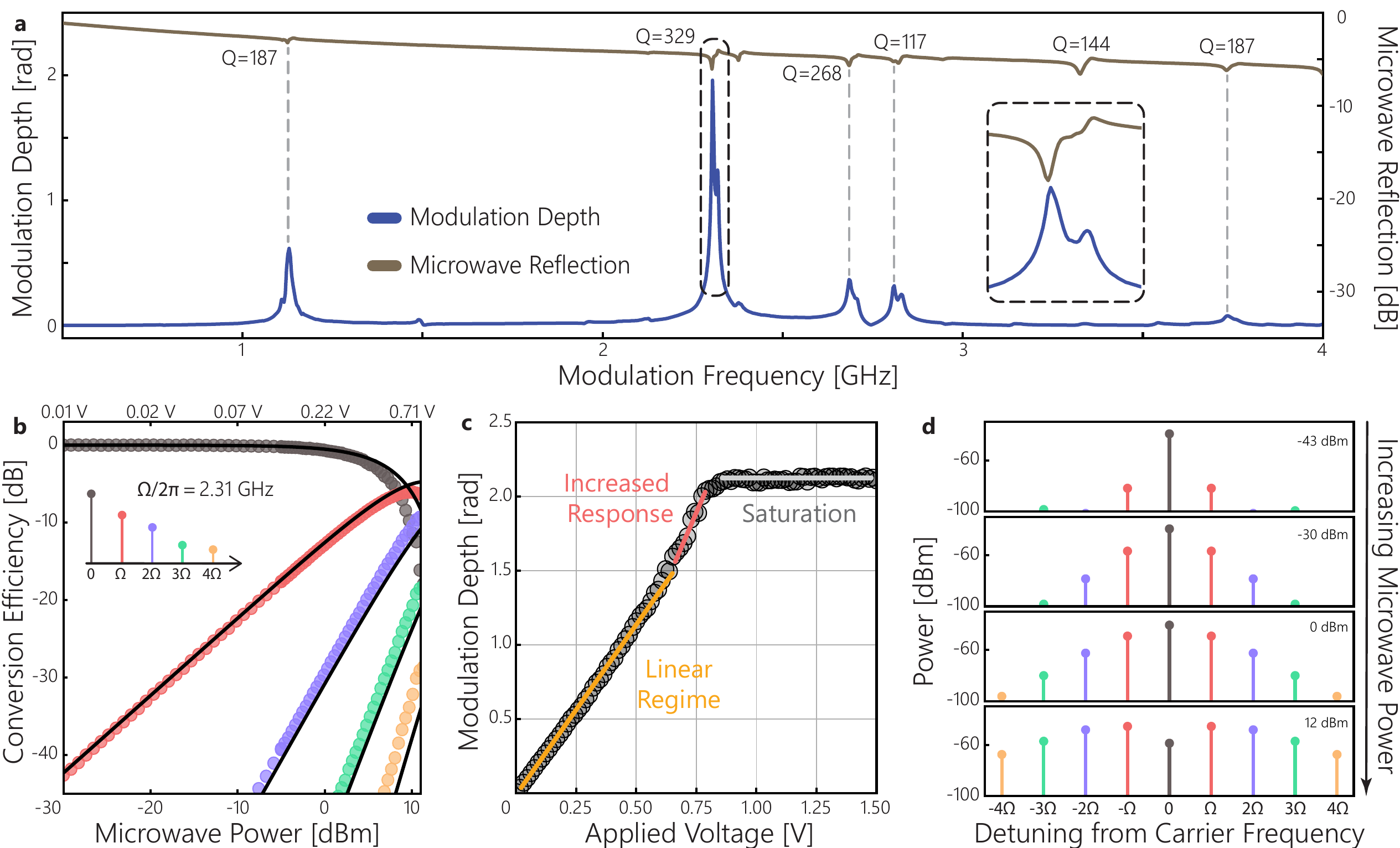}
    \caption{\label{fig:wide}{\bf Integrated acousto-optic phase modulation at visible wavelengths.} {\bf a} Plot showing modulation depth and microwave reflection as a function of frequency at a constant microwave drive power of $12 \text{ dBm}$. A strongly transduced mechanical resonance at $\SI{2.31}{\giga\hertz}$ produces a modulation depth of $\SI{2.1}{\radian}$, exceeding the point at which conversion to the first sideband reaches the theoretical maximum. The quality factors of prominent mechanical resonances are labeled adjacent to their corresponding dip in microwave reflection. {\bf b} Conversion efficiency to sidebands detuned from the laser frequency by integer multiples of $\Omega / 2\pi = \SI{2.31}{\giga\hertz}$ swept over applied microwave power and averaged over ten measurements, which displays close agreement with the Bessel functions predicted by the theory of sinusoidal phase modulation. {\bf c} On-resonance modulation depth as a function of drive power. The fitted line yields $V_{\pi} = \SI{1.32}{\volt}$ and $V_{\pi}\cdot L = \SI{0.26}{\volt\centi\metre}$. For higher drive voltages the phase modulation's growth saturates. {\bf d} Optical spectrum of light leaving the device after being phase modulated at $\Omega / 2\pi = \SI{2.31}{\giga\hertz}$ for various applied microwave powers, each averaged over ten measurements. Strong phase-modulation is evidenced by the cascaded transfer of optical power from each sideband to its neighbors.}
\end{figure*}

\begin{figure*}[htpb!]
    \includegraphics[width=17.5cm]{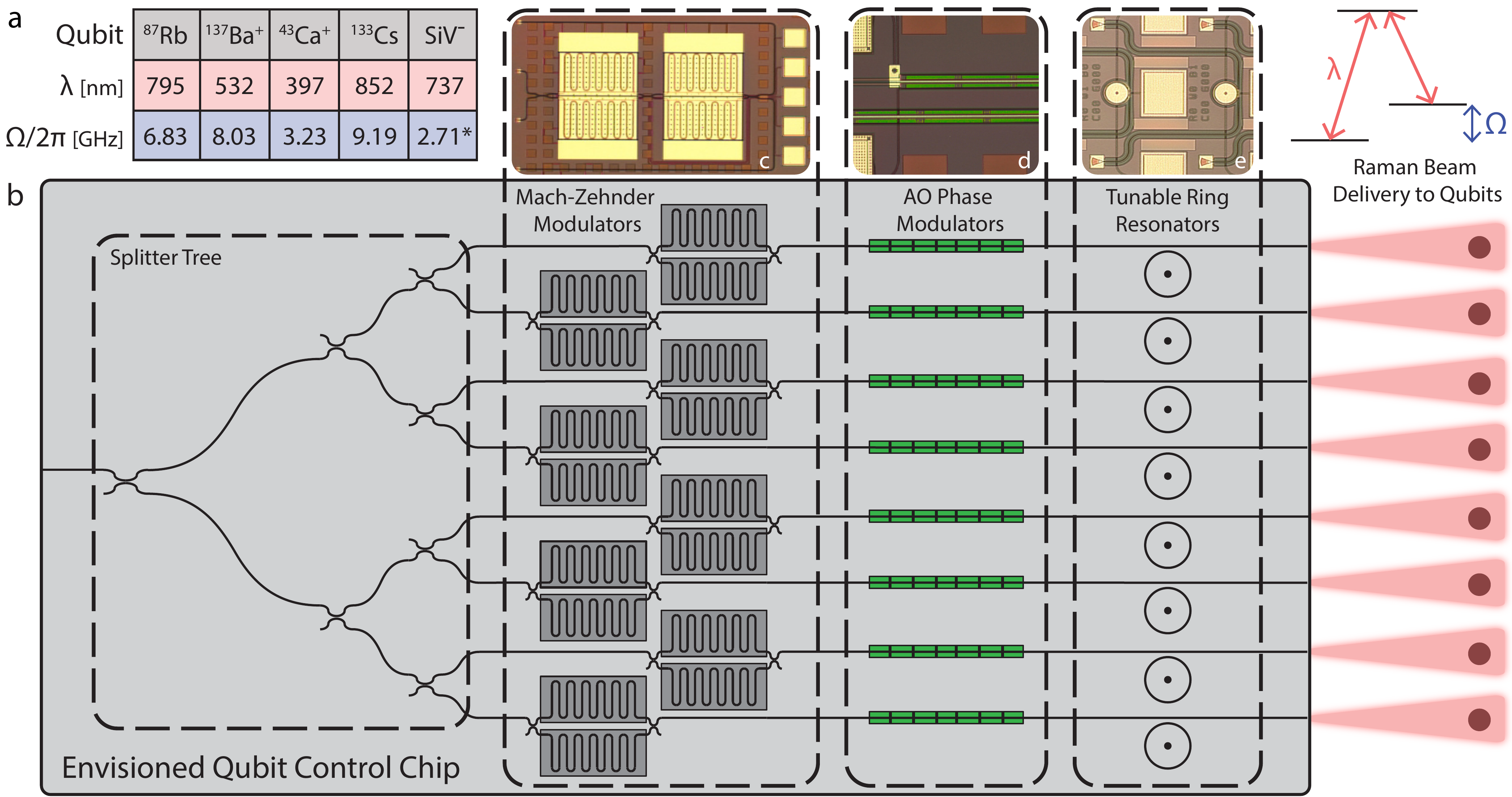}
    \caption{\label{fig:wide}{\bf Future platform integration for a qubit control chip.} {\bf a} Microwave frequencies and optical wavelengths associated with the transitions used for Raman control of typical qubit species \protect{\cite{park_technologies_2024}}. $^*$Dependent on externally applied magnetic field. {\bf b} Schematic of the envisioned qubit control chip. Light from a single input is split into many channels, each of which can be individually switched with a Mach-Zehnder modulator, frequency shifted with an acousto-optic phase modulator, and spectrally filtered with a tunable ring resonator. {\bf c} Micrographs of a Mach-Zehnder modulator \protect{\cite{dong_high-speed_2022, hogle_high-fidelity_2023}}, {\bf d} acousto-optic phase modulator, and {\bf e} tunable ring resonator \protect{\cite{stanfield_cmos-compatible_2019, menssen_scalable_2023}}, all of which are now demonstrated in our piezo-optomechanical photonic integrated circuit platform.}
\end{figure*}

\section{\label{sec:level1}EXPERIMENTAL RESULTS}
We characterize the fabricated devices with the setup shown schematically in Fig. 2c. A laser source is coupled to a fiber interferometer whose upper arm contains the chip on which the devices were fabricated and whose lower arm contains a commercially available acousto-optic frequency shifter (AOFS). In the upper arm, light from the fiber is grating-coupled to and from a waveguide that routes an optical mode through a particular phase modulator device. Meanwhile, light in the lower arm is passed through the AOFS which shifts its frequency by an amount $\Delta / 2\pi = \SI{125}{\mega\hertz}$. The arms are subsequently recombined on a $50/50$ directional coupler and directed to a fast photodetector whose electrical output is connected to a radio frequency spectrum analyzer (RFSA). 

We supply microwave power from a vector network analyzer (VNA) at an angular frequency, $\Omega$, to the electrodes of the on-chip device (by means of the electrical pads and interconnects imaged in Fig. 2b and graphically depicted in Fig. 1c). This excites a sinusoidal phase modulation process and produces the sidebands of equation \eqref{eq:jacobi}. These sidebands then interferometrically mix with the light of the lower arm to generate intensity modulation of the form $I(t) = I_0\sum_n J_n(\alpha)\cos\left[ (n\Omega + \Delta)t\right]$, where $I_0$ is a constant determined by the initial laser strength and the various loss mechanisms present in the setup ({\it e.g.} grating coupler insertion loss and waveguide propagation loss). An explanation for the AOFS's presence in the lower arm and a detailed derivation of the intensity signal are provided in Supplement IIIa. The microwave power at the angular frequencies $n\Omega + \Delta$ can then be recorded on the RFSA as a function of the driving frequency $\Omega$. Since these powers are proportional to $J_n^2(\alpha)$, they are proportional to the optical power produced by the on-chip phase modulator at the angular frequency $\omega - n\Omega$. The plot in Fig. 2a. shows these powers for $n = 0,1,2$ when driven at a constant microwave power of $\SI{15}{\milli\watt}$ for the device with a cladding width of $w_1 = \SI{1.25}{\micro\metre}$, core width of $w_2 = \SI{500}{\nano\metre}$, and a length of $L = \SI{2}{\milli\metre}$. We observe resonantly enhanced modulation at several frequencies including the one at $\SI{2.31}{\giga\hertz}$ at which the optical power in both the $n=1$ and $n=2$ sidebands surpass that of the remaining carrier, corresponding to appreciable modulation depths. 

To precisely determine the modulation depth, we extract it from the ratio of the measured optical power of two distinct sidebands at the microwave frequencies $m\Omega + \Delta$ and $n\Omega + \Delta$ (details of this method and its verification found in Supplement IIIb). The calculated modulation depth is plotted in Fig. 3a along with the microwave reflection coefficient $S_{11}$ measured by the VNA. Dips in the $S_{11}$ curve indicate those frequencies at which mechanical resonances are transduced, and they are labeled with their associated quality factors (which are calculated by a fit detailed in Supplement IIIc). As expected, the locations of these resonant mechanical dips closely agree with the frequencies at which peaks in modulation depth occur. We also observe that the regions of resonantly enhanced modulation depth consist of doublet peaks. This is because the device has discrete translational symmetry due to the periodic nanopillar supports along its length (as depicted in Fig. 1d). This makes the device a weak phononic crystal and gives rise to the doublets \cite{khelif_phononic_2015}. The presence of a strongly excited mechanical mode does not necessarily yield a large modulation depth because the degree to which a given mechanical mode induces changes to the optical refractive index is mediated by both the mechanical quality factor and the strength of the optomechanical coupling. This is evidenced by drive frequencies at which large dips in $S_{11}$ appear yet correspond only to a weak or nonexistent rise in modulation depth.

For the $\SI{2.31}{\giga\hertz}$ resonance, we observe a linear dependence between the modulation depth and voltage amplitude $V$ of the applied microwave signal until $V\approx \SI{0.65}{\volt}$, past which the modulation depth temporarily becomes more responsive before saturating completely at $V\approx \SI{0.85}{\volt}$ (Fig. 3c). We expect that a combination of thermal and mechanical nonlinearities produce this saturation, but the cause warrants further investigation \cite{mayor_gigahertz_2021}. Although a modulation depth of $\SI{2.1}{\radian}$ is sufficient for sideband generation, some applications benefit from stronger modulation \cite{karpinski_bandwidth_2017}, which future work could achieve by first identifying the precise cause of the saturation.  

Using the fitted $V_\pi$ to convert from drive power to modulation depth, we theoretically calculate the conversion efficiency to each of the sidebands of a pure phase modulation process and compare this against our experimentally obtained values in Fig. 3b. Excellent agreement is observed until the aforementioned nonlinearities cause the phase modulation process to begin accelerating at a microwave power of $8.00\text{ dBm}$ before ultimately saturating at $12.00\text{ dBm}$. Within this range of drive powers, $J_0(\alpha)$ is decreasing, $J_1(\alpha)$ reaches a maximum, and $J_{2,3,4}(\alpha)$ are increasing; this explains the theoretical overestimation of the $n=0$ and $1$ sidebands and underestimation of the rest.

We extract a mechanical quality factor of $Q=228$ from a fit to the modulation depth data at the $\SI{2.31}{\giga\hertz}$ resonance, implying a fast switching time of $\SI{31.4}{\nano\second}$. We also measure the on-resonance impedance of the device to be $\SI{149}{\ohm}$, which corresponds to $\SI{50}{\percent}$ microwave power-coupling. In future work impedance matching networks could be used to improve the microwave coupling efficiency \cite{pozar_microwave_2012}.

\section{\label{sec:level1}DISCUSSION}

We have demonstrated a visible-light, gigahertz-frequency acousto-optic modulator fabricated in a wafer-scale CMOS process. Our device produces resonantly enhanced modulation depths exceeding $\SI{2.1}{\radian}$ at a frequency of $\SI{2.31}{\giga\hertz}$ when driven with $\SI{15}{\milli\watt}$ of microwave power. These metrics indicate superior performance to state-of-the-art resonant, visible-light phase modulators with high power handling, which typically consume watt-level powers to produce modulation depths of order $\SI{1}{\radian}$ \cite{qubig_resonant_nodate, thorlabs_free_space}. 

Atomic and ionic species have strong optical transitions within the ultraviolet to near infrared and microwave transitions in the gigahertz range (selected species listed in Fig. 4a) \cite{bruzewicz_trapped-ion_2019, park_technologies_2024, harty_high-fidelity_2014}. Qubits are often encoded in the microwave levels, and gates are applied with Raman beams which have a carrier frequency detuned from the optical transition and exhibit amplitude modulation at the microwave transition frequency \cite{levine_dispersive_2022}. While the demonstration herein specifically takes place at $\SI{730}{\nano\metre}$ and $\SI{2.31}{\giga\hertz}$, the platform is straightforwardly extendable to a broad range of wavelengths and modulation frequencies that are utilized for Raman control. In Supplement V, we present simulations that show our modulators can operate at wavelengths spanning from $\SI{400}{\nano\metre}$ to $\SI{1000}{\nano\metre}$. We further show that by tuning the width of the device, strongly optomechanically coupled mechanical resonances are found over a frequency range spanning $\SI{1}{\giga\hertz}$ to $\SI{5}{\giga\hertz}$. In this range of operation, our acousto-optic phase modulator will enable the generation of gigahertz frequency shifts applicable to a variety of qubit species. By integrating this work with devices that have been demonstrated in the same piezo-optomechanical platform, we envision a photonic control chip that generates and delivers Raman beams to atomic and ionic qubits (shown schematically in Fig. 4b). The chip will fan out a single laser input to several channels, each of which can be individually amplitude modulated by Mach-Zehnder modulators with sub-microsecond switching times \cite{hogle_high-fidelity_2023, dong_high-speed_2022}, frequency modulated by the devices presented in this work, and spectrally filtered by tunable ring resonators with quality factors exceeding $1.5$ million \cite{stanfield_cmos-compatible_2019, menssen_scalable_2023}. The latter component is vital because the light incident on the qubits must be amplitude modulated at the qubit transition frequency. The pure phase-modulation process that generates the frequency modulation contributes zero amplitude modulation but can be converted to it by narrowband filtering of the carrier or sideband tones provided by the rings. Filtering the carrier tone produces amplitude modulation at twice the modulation frequency, which can be advantageous for qubits like $^{133}$Cs or $^{137}$Ba$^+$ that have relatively large hyperfine ground state splitting frequencies (see Fig. 4a). Additionally, the splitter tree and Mach-Zehnder modulator layers can be combined into a binary tree mesh that performs both switching and reconfigurable power routing, as was previously demonstrated in the same platform as the devices presented herein \cite{palm_modular_2023}.

By covering both the optical and microwave transitions of typical atomic qubits, the class of devices presented in this work is poised to uniquely enable large-scale, channelized frequency control for cooling, state preparation, gate operations, and readout. Although the transparency window of SiN$_x$ precludes operation deep into the ultraviolet, future work can utilize an alumina-based platform, in which we have already demonstrated piezo-optomechanical switches and tunable microresonators, to extend operation to this regime \cite{shugayev_alumina_2024}. We can further expand on our acousto-optics platform to leverage mechanical waves that copropagate with the optical mode to excite intermodal Brillouin scattering processes, enabling nonmagnetic isolation \cite{zhou_nonreciprocal_2024} and single-sideband frequency modulation. The latter of these capabilities is also immediately implementable using the phase modulators of this work arranged in a dual parallel Mach-Zehnder modulator configuration \cite{kodigala_high-performance_2024}.

\begin{acknowledgments}This material is based upon work supported by the U.S. Department of Energy, Office of Science, National Quantum Information Science Research Centers, Quantum Systems Accelerator. This work was performed, in part, at the Center for Integrated Nanotechnologies, an Office of Science User Facility operated for the U.S. Department of Energy (DOE) Office of Science. Sandia National Laboratories is a multimission laboratory managed and operated by National Technology \& Engineering Solutions of Sandia, LLC, a wholly owned subsidiary of Honeywell International, Inc., for the U.S. DOE's National Nuclear Security Administration under contract DE-NA-0003525. The views in the article do not necessarily represent the views of the U.S. Department of Energy or the United States Government.
\end{acknowledgments} 

\bibliography{references}

\end{document}


\begin{center}
    {\bf Supplement for \say{Gigahertz-Frequency Acousto-Optic Phase Modulation of Visible Light in a CMOS-Fabricated Photonic Circuit}}
\end{center}
\vspace{1em}
\begin{center} {\bf S-I: Effects of Phase-Matching.} \end{center}

For short device lengths, the phase of the microwave signal along the length of the device's electrode is approximately constant at given instant in time. Consequently, the phase of the excited mechanical deformations is also uniform along the device's length. Additionally, if the optical mode propagates through the device in a much shorter time than a period of the microwave signal, it will sample a single instantaneous phase of the mechanical deformation. Since the optical mode propagating through the modulator only samples a single instantaneous phase of the mechanical deformation along its traversal of the device, the phase shift, $\Delta\phi(t)$, it accumulates at time $t$ is then given by
\begin{align}
    \Delta\phi(t) &= \int_0^l \abs{\Delta\beta} \sin(\Omega t) dz
    \\[1.3em]
    \Delta\phi(t) &= \alpha\sin(\Omega t),
\end{align}
where $\Omega$ is the angular frequency of the driving voltage, $\abs{\Delta\beta}$ is the maximum change to the propagation constant of the optical mode induced by the mechanical deformation, $L$ is the length of the device, and $\alpha = \abs{\Delta\beta} \cdot L$. 

If the device length is made large, we must consider that the optical wavefront propagates a distance $z$ along the device in a non-negligible time given by $t = zn/c$, where $n$ is the effective refractive index of the mode \cite{scott_design_1995}. Taking this and the non-constant phase of the microwave signal with effective refractive index $N$ into account, the phase shift at time $t$ is now given by
    \begin{align}
        \Delta\phi(t) &= \int_0^L\abs{\Delta\beta} \sin\left[\Omega\left(t+\frac{nz}{c} - \frac{Nz}{c}\right)\right] dz
        \\[1.3em]
        \Delta\phi(t) &= \abs{\Delta\beta}\int_0^L \sin\left[\Omega\left(t-\frac{\Delta nz}{c}\right)\right] dz
        \\[1.3em]
        \Delta\phi(t) &= \abs{\Delta\beta} \frac{c}{\Omega\Delta n} \cos\left[\Omega\left(t - \frac{\Delta n z}{c}\right)\right]_0^L
        \\[1.3em]
        \Delta\phi(t) &= \frac{c\abs{\Delta\beta}}{\Omega\Delta n} \left[ \cos\left(\Omega t-\frac{\Omega\Delta n L}{c}\right) - \cos\left(\Omega t\right) \right]
        \\[1.3em]
        \Delta\phi(t) &= \frac{2c\abs{\Delta\beta}}{\Omega\Delta n}\sin\left( \frac{\Omega\Delta nL}{2c} \right)\sin \left( \Omega t - \frac{\Omega\Delta n L}{2c} \right)
        \\[1.3em]
        \Delta\phi(t) &= \abs{\Delta\beta}L\cdot\sinc\left( \frac{\Omega\Delta nL}{2c} \right)\sin \left( \Omega t - \frac{\Omega\Delta n L}{2c} \right).
    \end{align}
    We can see that the effect of the difference in propagation speeds (quantified by $\Delta n = n - N$) is to scale the modulation depth by a factor of $\sinc\left( \frac{\Omega\Delta nL}{2c} \right)$ from its phase-matched value and to shift the overall phase of the modulation. Thus in order to operate in a regime that ignores the effects of phase-matching, we must have $L \ll 2c / \Omega\Delta n$. For a conservative estimate of $\Delta n \approx 2$ and operating at $\Omega/2\pi = \SI{2}{\giga\hertz}$, this yields $L \ll \SI{2.4}{\centi\metre}$. The device considered in the main text is $\SI{2}{\milli\metre}$ long, which satisfies this condition. 
\\
\newpage 
\begin{center} {\bf S-II: Optomechanical and Electromechanical Coupling.} \end{center}

\begin{figure}[H]
    \centering
    \includegraphics[width=1\linewidth]{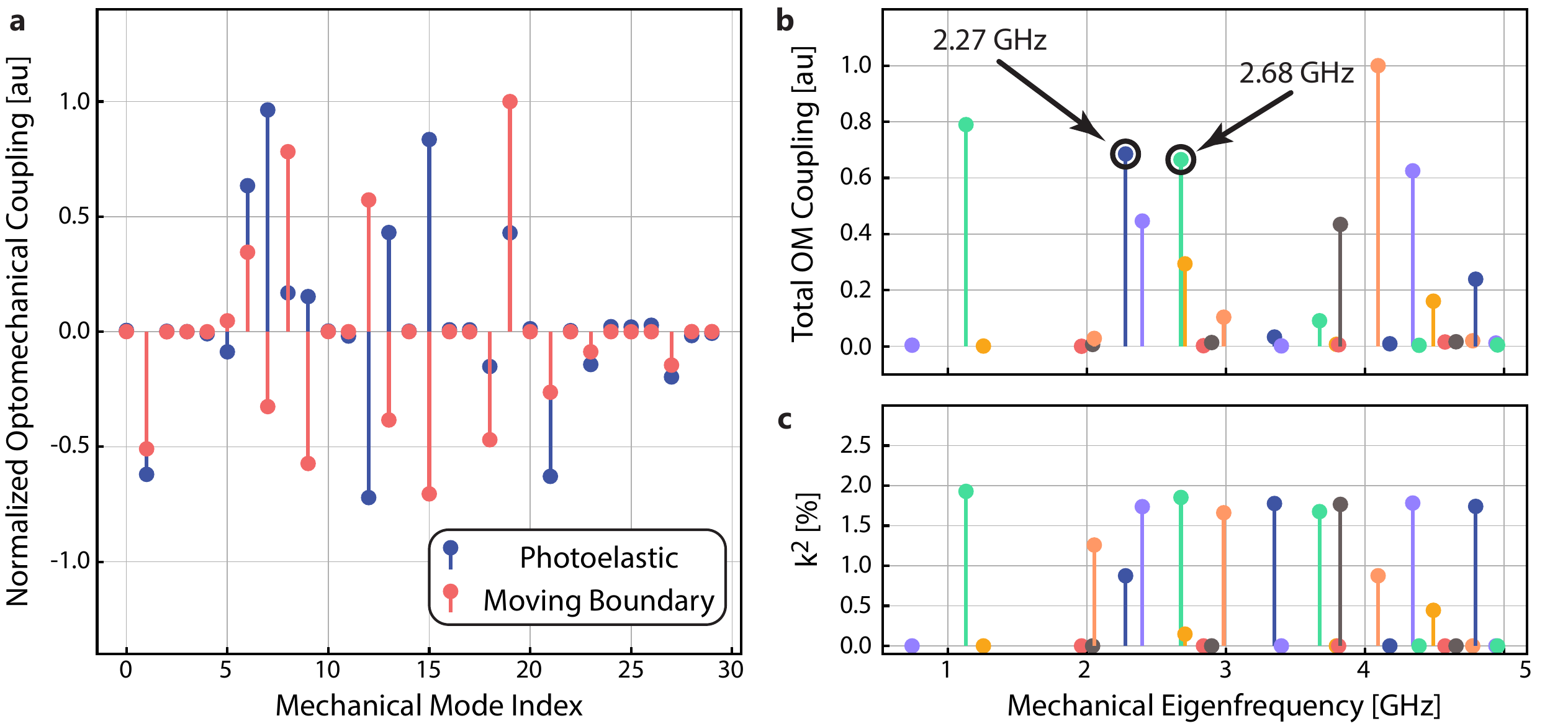}
    \caption{{\bf Optomechanical coupling simulations.} {\bf a} Simulated normalized photoelastic and moving boundary optomechanical coupling to the fundamental TE optical mode for the first $30$ mechanical breathing modes. {\bf b} Total optomechanical coupling for the first $30$ mechanical breathing modes to the fundamental TE optical mode plotted as a function of the mechanical modes' eigenfrequencies. {\bf c} Plot of the electromechanical coupling coefficient, $k^2$, as a function of the mechanical modes' eigenfrequencies. Colors are cycled for ease of comparison with the plot in {\bf b}.}
    \label{fig:omc}
\end{figure}

Mechanical deformations within an optical waveguide modify the optical mode's propagation constant through two effects: the photoelastic effect and the moving boundary effect \cite{wiederhecker_brillouin_2019}. The individual contributions of these effects are calculated by perturbation theory \cite{johnson_perturbation_2002} as
\begin{align}
    \Delta \beta_{\text{pe}} &= \frac{1}{d\omega / d\beta} \frac{\omega}{2} \frac{\iint\mathbf{E}^*\cdot\Delta\mathbf{\epsilon}\cdot\mathbf{E}\phantom{x}dA}{\max{|\mathbf{u}|}\iint \mathbf{E}^*\cdot\mathbf{\epsilon}\cdot\mathbf{E}\phantom{x}dA},
     \label{eq:pe} \\[1.5em]
    \Delta \beta_{\text{mb}} &= \frac{1}{d\omega / d\beta} \frac{\omega}{2} \frac{\oint \mathbf{u}
    \cdot\hat{\mathbf{n}}\left[\Delta\epsilon_{12}(\mathbf{E}_{||}^*\cdot\mathbf{E}_{||})-\Delta\epsilon_{12}^{-1}(\mathbf{D}_{\perp}^{*}\cdot\mathbf{D}_\perp)\right]dl}{\max{|\mathbf{u}|}\int \mathbf{E}^*\cdot\mathbf{\epsilon}\cdot\mathbf{E}\phantom{x}dA} \label{eq:mb}.
\end{align}
The photoelastic integral of equation \eqref{eq:pe} is evaluated over a plane perpendicular to the axis along which the optical mode propagates, and $\omega$ is the optical mode's angular frequency, $\mathbf{E}$ is the optical mode's electric field profile, $\mathbf{u}$ is the mechanical displacement field, $\epsilon$ is the material permittivity tensor, $\Delta\epsilon$ is the permittivity tensor perturbation photoelastically induced by the strain field, and $d\omega / d\beta$ is the optical mode's group velocity. The moving boundary integral of equation \eqref{eq:mb} is evaluated along the boundary between the SiN$_x$ core and SiO$_2$ cladding, and $\hat{\mathbf{n}}$ is the outward-directed normal vector to the boundary, $\mathbf{E}_{||}$ is the optical mode's electric field parallel to the boundary, $\mathbf{D}_\perp$ is the optical mode's electric displacement field perpendicular to the boundary, $\Delta\epsilon_{12}$ is the difference in permittivity between the core and cladding, and $\Delta\epsilon_{12}^{-1}$ is the difference in the reciprocal of permittivity between the core and cladding. For a given device geometry (Fig. 1a), we simulated the breathing mechanical resonances and propagating optical modes of the structure with finite element method (FEM) software (Fig. 1b-c). From these simulations, we obtain all of the relevant fields and parameters present in equations \eqref{eq:pe} and \eqref{eq:mb}. The photoelastic permittivity perturbation is calculated in Voigt notation as
\begin{align}
    \frac{1}{\Delta\epsilon_i} = \sum\limits_{j=1}^{6} p_{ij}S_j,     
\end{align}
where $p_{ij}$ is the photoelastic tensor and $S_j$ is the strain tensor \cite{tian_piezoelectric_2024}. The two materials, SiO$_2$ and SiN$_x$, that comprise the optical waveguide are each isotropic and therefore have photoelastic tensors of the form
\begin{align}
    p_{ij} = \begin{bmatrix} 
                    p_{11} & p_{12} & p_{12} & 0 & 0 & 0 \\
                      p_{12} & p_{11} & p_{12} & 0 & 0 & 0 \\
                      p_{12} & p_{12} & p_{11} & 0 & 0 & 0 \\
                      0 & 0 & 0 & p_{44} & 0 & 0 \\
                      0 & 0 & 0 & 0 & p_{44} & 0 \\
                      0 & 0 & 0 & 0 & 0 & p_{44} \end{bmatrix},
\end{align} 
where $p_{44} = (p_{11} - p_{12})/2$. Since our device operates with transverse electric optical modes, we are primarily concerned with the first row of the photoelastic tensor, which is responsible for modifications to the $\epsilon_1$ (or $\epsilon_{xx}$) component of the permittivity tensor. Even so, our calculations of optomechanical coupling consider the tensor in its entirety for completeness. With this, we numerically evaluate the integrals for simulated mechanical-optical mode pairings. Normalized photoelastic and moving boundary couplings are shown in Fig. \ref{fig:omc}a between 30 different mechanical breathing modes and the fundamental transverse electric optical mode. Despite the moderate index contrast between SiN$_x$ and SiO$_2$ of $0.55$, we find that the moving boundary effect's contribution is generally of the same order as that of the photoelastic effect, and even sometimes exceeds it. The respective perturbations to the propagation constant from the photoelastic and moving boundary effects can only differ in phase by zero or $\pi$ because the permittivity perturbation, $\Delta\epsilon$, in \eqref{eq:pe} includes the phase of the mechanical strain field, and the mechanical displacement vector, $\mathbf{u}$, in \eqref{eq:mb} also includes that same phase (because the components of strain are linear combinations of the derivatives of the components of displacement). Thus, as in Fig. S1a, we can remove a factor of constant phase and consider $\Delta\beta_{\text{pe}}$ and $\Delta\beta_{\text{mb}}$ as real-valued quantities which either combine or counteract one another. In Fig. \ref{fig:omc}b, we plot the absolute value of the total optomechanical coupling (the sum of the photoelastic and moving boundary couplings) against the frequency of the mechanical modes.

We also calculate the electromechanical coupling coefficient $k^2$ for each mechanical mode by performing FEM simulations, in which the electrodes of the device are driven with an oscillating voltage. From these simulations we extract the device's series resonance frequency, $\omega_s$, and parallel resonance frequency, $\omega_p$, associated with each of the mechanical modes simulated in Fig. \ref{fig:omc}a \cite{sanchez-rojas_piezoelectric_2020}. We then obtain the coupling coefficient from
\begin{align}
    k^2 = \frac{\omega_{p}^2 - \omega_s^2}{\omega_p^2}.
\end{align}

The values of $k^2$ are plotted in Fig. \ref{fig:omc}c along the same frequency axis as the values of optomechanical coupling in Fig. \ref{fig:omc}b. In general, modes with significant optomechanical coupling also have nonzero $k^2$. This is because the same mechanical mode symmetries that allow for the integrals of equations \eqref{eq:pe} and \eqref{eq:mb} to be nonzero enable the electric field driven by the device's electrodes to effectively overlap with the piezoelectrically generated electric field associated with the mechanical mode. We find that the simulated mode at $\SI{2.68}{\giga\hertz}$ labeled in black in Fig. \ref{fig:omc}b has both high optomechanical and electromechanical coupling. Yet, experimentally we find the mode at $\SI{2.31}{\giga\hertz}$, which we believe corresponds to the simulated mode at $\SI{2.27}{\giga\hertz}$, produces the most efficient modulation. This discrepancy and similar inconsistencies can be explained by the much larger quality factor we measure for the $\SI{2.31}{\giga\hertz}$ resonance (see S-IIIc for quality factor details).

\begin{center} {\bf S-III: Measurement Techniques.} \end{center}

\noindent {\bf S-IIIa: Calculation of Intensity Signal.} Consider that the optical signal from the chip in the upper arm of the interferometer and entering the photodetector of Fig. 2c has the form $E_u(t) = A_u\exp\left\{i\left[-\omega t + \alpha\sin\left(\Omega t\right)\right]\right\}$, and similarly that of the lower arm has the form $E_l(t) = A_l \exp\left\{i\left[-\left(\omega+\Delta\right) t - \phi \right]\right\}$, where $A_u$ and $A_l$ are the real-valued amplitudes of the fields in the upper and lower arm, respectively, $\omega$ is the laser angular frequency, $\alpha$ is the modulation depth, $\Omega$ is the angular frequency of the microwave drive, $\Delta$ is angular frequency shift imparted by the commercial AOFS, and $\phi$ is the static phase-difference between the two fields due to the unbalanced state of the interferometer, which also accounts for the phases imparted by the 2x2 fiber directional couplers. We assume that each frequency component generated by the phase-modulation process is subject to the same transmission coefficient associated with the various components in the measurement setup, including the grating couplers, fiber directional couplers, and any Fabry-Perot cavities formed between chip or fiber facets. Since the modulation depths considered here are small enough that the vast majority of the optical power will remain within the first few sidebands, the assumption is valid barring the presence of any extremely dispersive or frequency selective elements. With this we can write the intensity signal incident on the fast photodetector as

    \begin{align}
        I(t) &= \left[E_u(t) + E_l(t)\right]\left[E_u^*(t) + E_l^*(t)\right] \vphantom{\left\{\sum_{n=-\infty}^\infty\right\}}
        \\
        I(t) &= A_u^2 + A_l^2 + A_lA_u \left\{ \exp\left\{ i\left[ \Delta t + \alpha\sin\left(\Omega t\right) + \phi\right]\right\} + \exp\left\{ -i\left[\Delta t - \alpha\sin\left(-\Omega t\right) + \phi\right]\right\}\right\} \vphantom{\left\{\sum_{n=-\infty}^\infty\right\}}
        \\
        I(t) &= A_u^2 + A_l^2 + A_lA_u \left\{ \exp\left[i\left(\Delta t+\phi\right)\right]\sum_{n=-\infty}^\infty J_n (\alpha)\exp\left(in\Omega t\right) + \exp\left[-i\left(\Delta t+\phi\right)\right]\sum_{n=-\infty}^\infty J_n (-\alpha)\exp\left(in\Omega t\right)\right\}
        \\
        I(t) &= A_u^2 + A_l^2 + A_lA_u \left\{ \sum_{n=-\infty}^\infty J_n (\alpha)\exp\left\{i\left[\left(n\Omega+\Delta\right) t + \phi \right]\right\} + \sum_{n=-\infty}^\infty J_n (\alpha)\exp\left\{-i\left[\left(n\Omega+\Delta\right) t + \phi \right]\right\} \right\}
        \\
        I(t) &=  A_u^2 + A_l^2 + 2A_lA_u\sum_{n=-\infty}^\infty J_n(\alpha) \cos\left[\left(n\Omega + \Delta\right)t + \phi\right]. \vphantom{\left\{\sum_{n=-\infty}^\infty\right\}} \label{eq:signal}
    \end{align}
    We can see that the intensity signal consists of a constant term and a sum of terms weighted by Bessel functions that each oscillate at a distinct frequency, $n\Omega+\Delta$. In order to elucidate the importance of the AOFS in the lower arm, we will now derive the intensity $I^\prime$ in the case that $\Delta \rightarrow 0$; {\it i.e.} when there is no AOFS. We have that
    \begin{align}
        I^\prime(t) &= A_u^2 + A_l^2 + 2A_lA_u\sum_{n=-\infty}^\infty J_n(\alpha) \cos\left(n\Omega t + \phi\right). \vphantom{\left\{\sum_{n=-\infty}^\infty\right\}}
        \\
        I^\prime(t) &= A_u^2 + A_l^2 + 2A_lA_u \left[\cos\left(\phi\right)\sum_{n=-\infty}^\infty J_n(\alpha)\cos\left(n\Omega t\right) - \sin\left(\phi\right)\sum_{n=-\infty}^\infty J_n(\alpha)\sin\left(n\Omega t\right) \right]
        \\
        I^\prime(t) &= A_u^2 + A_l^2 + 4A_lA_u \left\{\cos\left(\phi\right)  \sum_{n=0}^\infty J_{2n}(\alpha) \cos\left(2n\Omega t\right) + \sin\left(\phi\right)  \sum_{n=1}^\infty J_{2n-1}(\alpha) \sin\left[\left(2n-1\right)\Omega t\right] \right\}
        \\
        I^\prime(t) &= A_u^2 + A_l^2 + 2A_lA_uJ_0(\alpha)\cos(\phi) + 4A_lA_u \left\{\cos\left(\phi\right)  \sum_{n=1}^\infty J_{2n}(\alpha) \cos\left(2n\Omega t\right) + \sin\left(\phi\right)  \sum_{n=1}^\infty J_{2n-1}(\alpha) \sin\left[\left(2n-1\right)\Omega t\right] \right\},
    \end{align}
    where we have made use of a trigonometric identity and the fact that for even $n$, $J_n(\alpha) = J_{-n}(\alpha)$, while for odd $n$, $J_n(\alpha) = -J_{-n}(\alpha)$. One salient difference between $I(t)$ and $I^\prime(t)$ is for the former the static phase difference $\phi$ uniformly shifts the phase of the oscillating sinusoids, while for the latter it modulates the intensity of the even and odd harmonics of $\Omega$ present at the photodetector. Thus, if there are any phase fluctuations in the fiber causing $\phi \rightarrow \phi(t)$ to become a random function of time, the readout of the strength of the intensities at the various frequencies $n\Omega$ will vary randomly in time; what is worse, since the intensities are modulated by $\cos\phi$ and $\sin\phi$, the average intensity of any given sideband will be zero given $\phi(t)$ covers a whole cycle of phase. The presence of the AOFS removes the issue of noise in the strength of the oscillating intensity terms and eliminates the need to phase-lock the interferometer. Also, we can see that $I^\prime(t)$ does not distinguish between $n$ and $-n$, whereas $I(t)$ maps them to distinct frequencies whose Fourier decomposition may be extracted by a spectrum analyzer. Although a pure phase-modulation process generates symmetric power conversion to the red and blue-shifted sidebands, the ability to distinguish them allows us to identify the presence of other processes such as deliterious frequency-discriminating Fabry-Perot cavities formed between fiber and waveguide facets or intermodal Brillouin scattering.
    \\

\begin{figure}[H]
    \centering
    \includegraphics[width=1\linewidth]{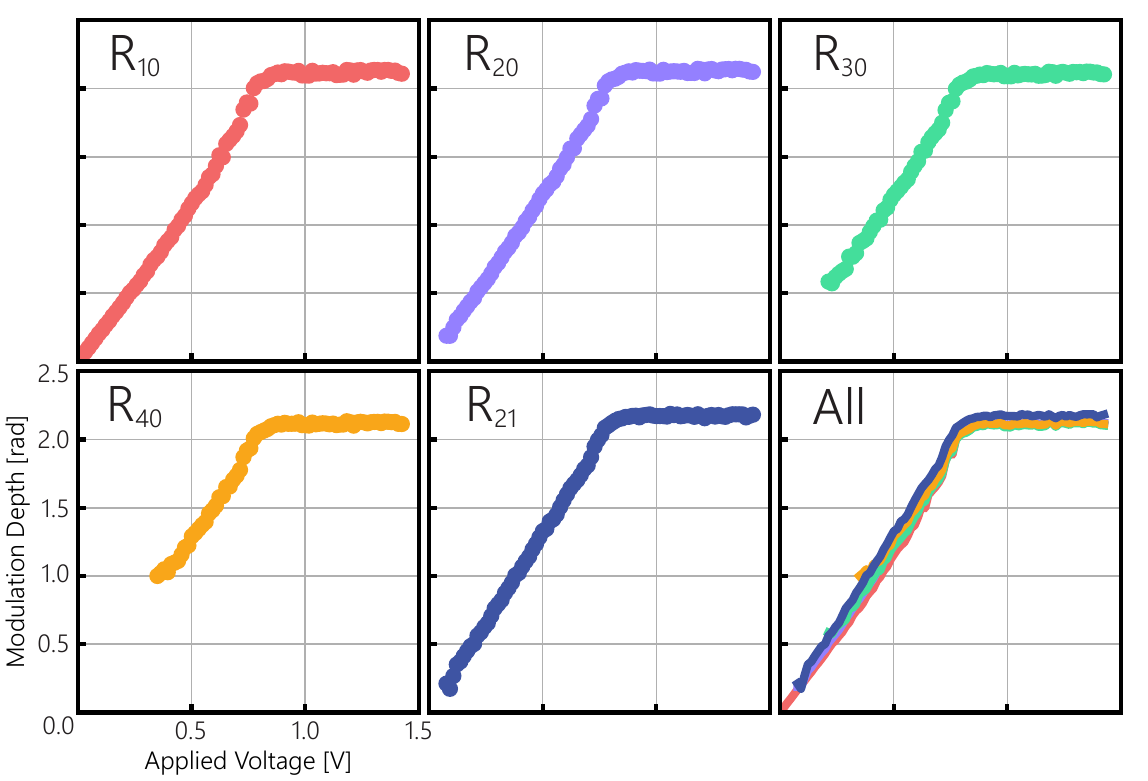}
    \caption{{\bf Comparison of measured modulation depths.} Modulation depths extracted from five different sideband power ratios $R_{mn}$. Driving voltage was swept at a fixed frequency of $\SI{2.31}{\giga\hertz}$, and the axes for each subplot are the same.}
    \label{fig:order_comp}
    \end{figure}

    \noindent {\bf S-IIIb: Determination of modulation depth.} The photodetector in our setup outputs the signal mathematically represented in equation \eqref{eq:signal}. We obtain the RF power $P_n$ at a given frequency $n\Omega+\Delta$ by passing this signal to a radio frequency spectrum analyzer (RFSA). For each $n$, these powers are proportional to $J_n^2(\alpha)$ with the same constant of proportionality, so $R_{nm} \overset{\text{def}}{=} P_n / P_m = J_n^2(\alpha) / J_m^2(\alpha)$. Since $R_{mn}$ is a value directly measured from the RFSA, and $J_n$ and $J_m$ are known functions, we can find a modulation depth $\alpha$ that satisfies this equation. To confirm the validity of these extracted modulation depths, we calculated and compared the $\alpha$ measured from several different sideband ratios $R_{mn}$. Fig. \ref{fig:order_comp} shows the measured $\alpha$ from five different sideband ratios as a function of driving voltage for the device reported in the main text, and we observe close agreement amongst them. For modulation depths calculated from ratios with higher order sidebands ({\it e.g.} $R_{40}$) the power signal on the RFSA does not emerge from the noise floor until a certain cutoff voltage, and the essentially random modulation depth measurements until that point are excluded from the plots.

\noindent {\bf S-IIIc: Determination of Mechanical Quality Factors.} We extract the mechanical quality factors of the observed mechanical modes from modulation depth versus frequency data. The photoelastic and moving-boundary perturbations to the optical mode's effective refractive index are proportional to the mechanical strain magnitude. Thus, the device's modulation depth is also proportional to the strain magnitude. Around nearly every frequency at which we observe resonant modulation, we find doublet mechanical resonances. Because the device has discrete translational symmetry from the periodic nanopillar supports, it behaves as a weak phononic crystal, which gives rise to the doublets.

\begin{figure}[H]
    \centering
    \includegraphics[width=1\linewidth]{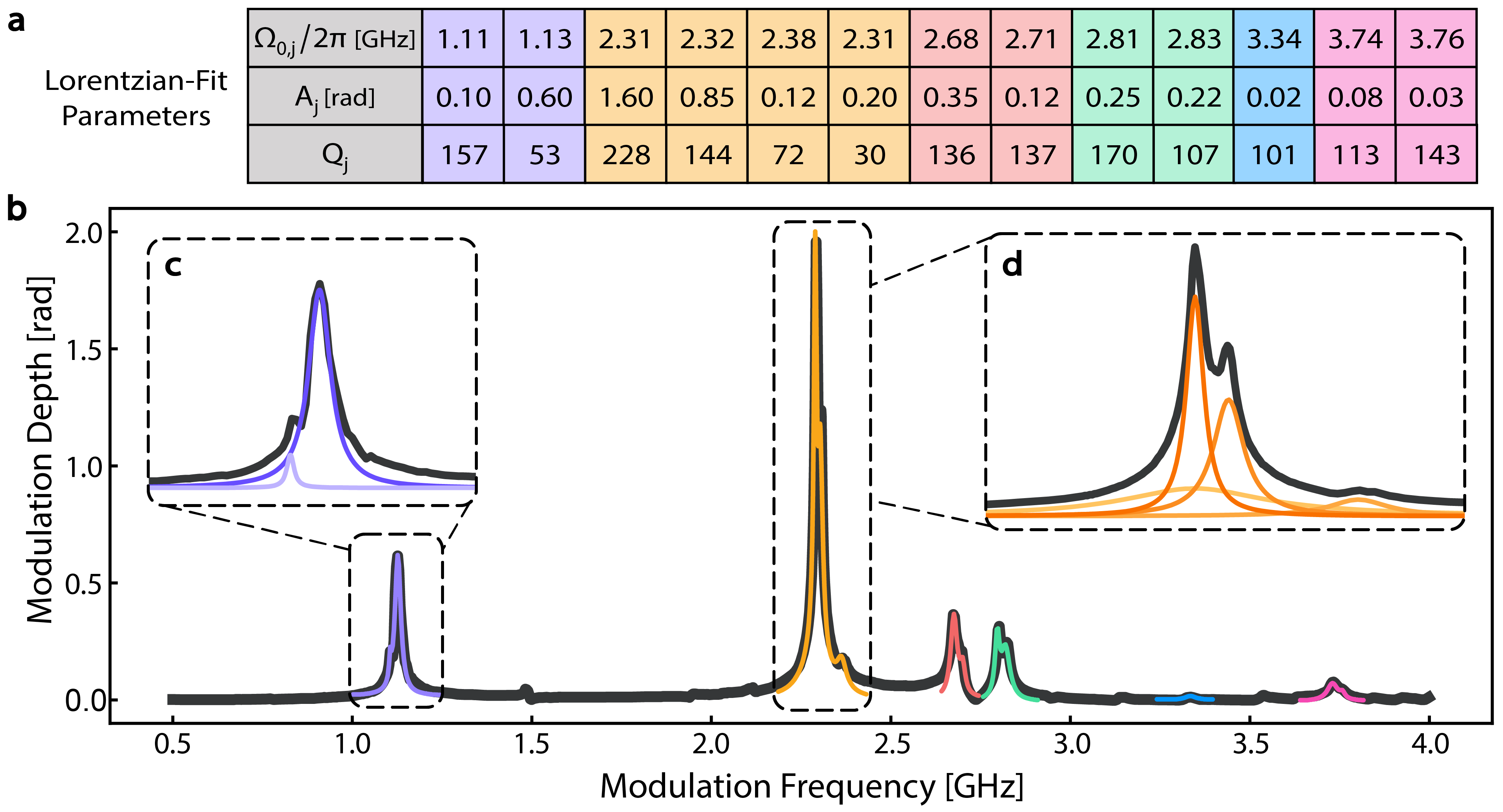}
    \caption{{\bf Determination of mechanical quality factors.} {\bf a} Lorentzian fit parameters including the central frequencies ($\Omega_{0,j} / 2\pi$), amplitudes $A_j$, and quality factors $Q_j$ for each resonance. Colors correspond to parameters fit from the same batch of data. {\bf b} Plot of the numerical fits (in color) superimposed on the measured modulation depth (in gray). Insets show in {\bf c} the Lorentzian doublet fit around $\SI{1.1}{\giga\hertz}$ and in {\bf d} the Lorentzian quadruplet fit around $\SI{2.3}{\giga\hertz}$.}
    \label{fig:qfactor}
\end{figure}

We numerically fit the measured modulation depth data $\alpha(\Omega)$ to a sum of Lorentzian functions of the form
\begin{align}
    \tilde{\alpha}(\Omega) = \sum\limits_j A_j \frac{\left(\Omega_{0,j}/2Q_j\right)^2}{\left(\Omega - \Omega_{0,j}\right)^2 + \left(\Omega_{0,j}/2Q_j\right)^2},
\end{align}
where $\Omega_{0,j}$ is $j$th resonance angular frequency, $A_j$ is the $j$th Lorentzian amplitude in units of radians, and $Q_j$ is the $j$th mechanical quality factor. The fitted parameters are listed in Fig. \ref{fig:qfactor}a and plotted in Fig. \ref{fig:qfactor}b-d. Fits are obtained from batches of data within the neighborhood of observed resonances. For the frequency range around $\SI{2.31}{\giga\hertz}$, we fit a sum of four Lorentzians (Fig. \ref{fig:qfactor}d) because two pairs of doublets existed within each other's bandwidths. For the resonance with central frequency $\SI{3.34}{\giga\hertz}$, we fit a single Lorentzian because no doublet-feature was observed. All other Lorentzians were fit in pairs, and parameters that were simultaneously extracted from the same batch of data are grouped by color in Fig. \ref{fig:qfactor}a.

\begin{center} {\bf S-IV: Simulation of Broadband Operation} \end{center}

\begin{figure}[H]
    \centering
    \includegraphics[width=1\linewidth]{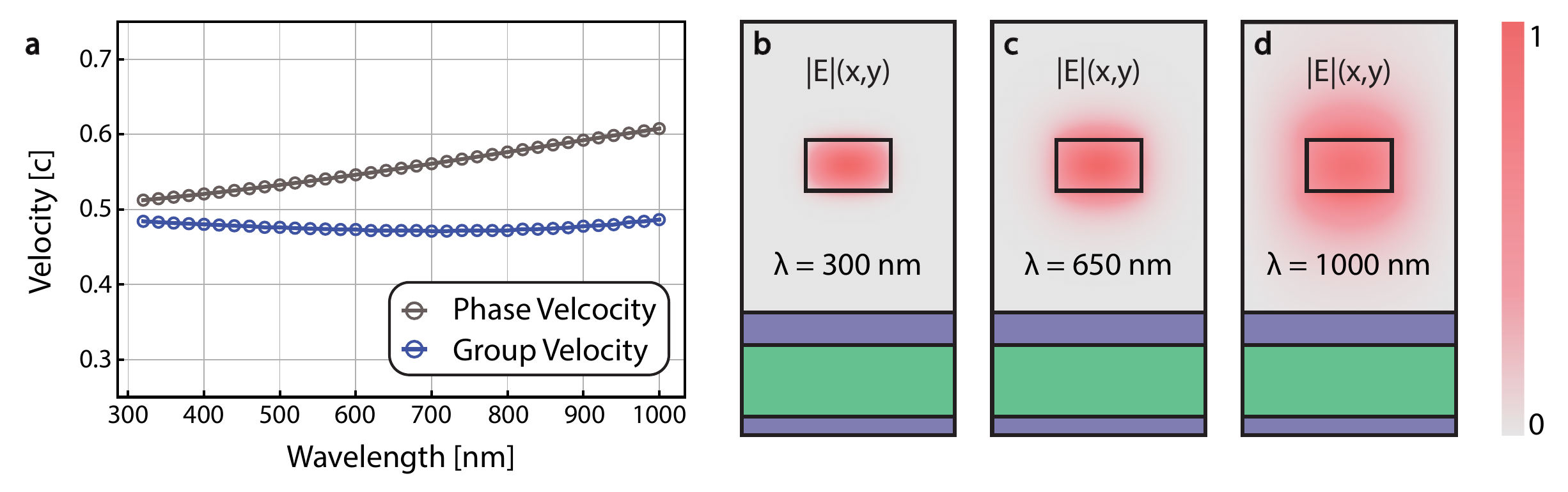}
    \caption{{\bf Optical mode as a function of wavelength.} {\bf a} Plot of group velocity and phase velocity in units of the free-space speed of light for the device's optical waveguide as a function of free-space wavelength. {\bf b-d} Plots of the electric field norm for the fundamental transverse electric mode at $\SI{300}{\nano\metre}$, $\SI{650}{\nano\metre}$, and $\SI{1000}{\nano\metre}$. The simulations from which these are derived include a large domain of air surrounding the waveguide to ensure the expanding optical mode does not interact with nonphysical boundary conditions.}
    \label{fig:group}
\end{figure}

We simulate the optical bandwidth of our modulator by calculating the photoelastic and moving boundary coupling of equations \eqref{eq:pe} and \eqref{eq:mb}, respectively, as a function of optical wavelength. We do so by performing FEM simulations of the structure from the main text for wavelengths spanning from $\SI{300}{\nano\metre}$ to $\SI{1000}{\nano\metre}$ in increments of $\SI{20}{\nano\metre}$. At a fixed optical wavelength, the group velocity parameter in the coupling equations does not play a role in the comparison of optomechanical coupling from distinct mechanical modes. Here, we are comparing coupling across wavelengths, so we calculate it from the simulations and plot it along with the mode's phase velocity in Fig. \ref{fig:group}a. We expectedly observe that the optical mode profile becomes decreasingly confined as wavelength is increased (Fig. \ref{fig:group}b-d). If the sign of the mechanical strain field changes sign near the SiN$_x$-SiO$_2$ boundary, a lack of confinement causes the mode to sample both negative and positive changes to refractive index, thereby decreasing the photoelastic coupling integral. Likewise, if the strain-field maintains the same sign near the waveguide boundary, decreased confinement can lead to increased photoelastic coupling. Moving boundary coupling is modified by the degree to which the optical mode is concentrated at the SiN$_x$-SiO$_2$ boundary, which is affected by the confinement of the mode. Despite the photoelastic and moving boundary effects having the above dependencies on wavelength, we expect the overall coupling to remain relatively constant over wavelength ranges spanning hundreds of nanometers because the modal profile varies slowly with respect to wavelength.

\begin{figure}[H]
    \centering
    \includegraphics[width=1\linewidth]{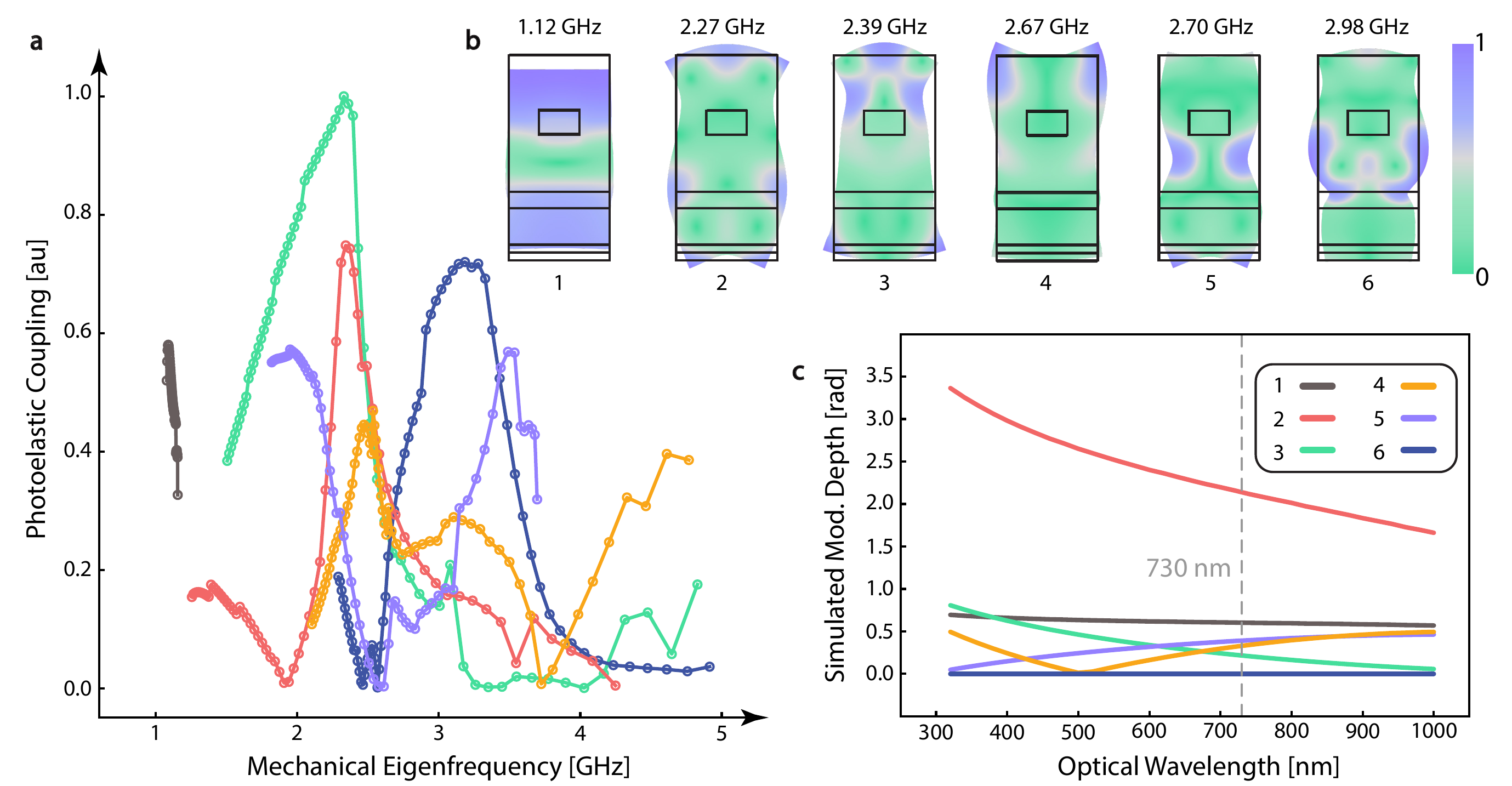}
    \caption{{\bf Simulated broadband operation of the phase-modulator.} {\bf a} Plot of normalized photoelastic optomechanical coupling against eigenfrequency for six mechanical resonances which are simulated as the device width varies from $\SI{600}{\nano\metre}$ to $\SI{2.3}{\micro\metre}$. {\bf b} The mechanical displacement profiles and eigenfrequencies of the six mechanical modes at a device width of $w_1=\SI{1.25}{\micro\metre}$. {\bf c} Plot showing simulated modulated depth as a function of optical wavelength for the same six mechanical modes as in {\bf a}. The simulated drive power was $\SI{15}{\milli\watt}$, and the modulation depths are calibrated to the experimentally measured values at $\SI{730}{\nano\metre}$. }
    \label{fig:bandwidth}
\end{figure}

Since predictions for modulation efficiencies calculated from simulated optomechanical coupling do not account for mechanical quality factor or device impedance, we calibrate out prediction from our experimental results. From the experimentally obtained data, we have the frequencies at which mechanical resonances occur and the modulation depth $\alpha_0^{(j)}$ the $j$th mode imparts at $\lambda_0 = \SI{730}{\nano\metre}$. By combining this information with simulations of the perturbation to the propagation constant $\Delta \beta^{(j)}(\lambda)$ imparted by the $j$th mechanical to the optical mode with free-space wavelength $\lambda$, we obtain a prediction for the modulation depth $\alpha(\lambda)$. We have that
\begin{align}
    \frac{\alpha^{(j)}(\lambda)}{\alpha^{(j)}_0} = \Delta \beta^{(j)}(\lambda)L \cdot \frac{1}{\Delta \beta^{(j)}(\lambda_0)L} = \frac{\Delta \beta^{(j)}(\lambda)}{\Delta \beta^{(j)}(\lambda_0)} \overset{\text{def}}{=} R^{(j)}(\lambda),
\end{align}
where $L$ is the length of the device. Thus, for the $j$th mechanical mode and some optical wavelength $\lambda$, we can predict the device's modulation depth to be $\alpha^{(j)}(\lambda) = R^{(j)}(\lambda)\alpha^{(j)}_0$. This calibration of the simulated prediction to the experimentally measured values at $\SI{730}{\nano\metre}$ has the effects of electromechanical coupling, mechanical quality factor, and microwave reflection baked into the values. Fig. \ref{fig:bandwidth}c shows these simulated modulation depths for three selected mechanical modes over the transparency window of the SiN$_x$ waveguide. From this, we can see that the modulator would function well over an optical bandwidth spanning hundreds of nanometers. As wavelength is decreased, the optical mode's rate of phase accumulation inherently increases because of the increase to its propagation constant. If $\Delta \beta^{(j)}(\lambda)$ is relatively constant as a function of wavelength, this increase in modulation depth toward shorter wavelengths is observed ({\it e.g.} for modes 2 and 3 in Fig. \ref{fig:bandwidth}c). However, for Mode 1 the optomechanical coupling strength decreases as a function of optical wavelength and counteracts the effect of increasing propagation constant, rendering its modulation depth nearly constant over the simulated wavelength range. 

For the Raman control of hyperfine qubits, our device must be capable of producing resonantly enhanced phase modulation at frequencies specific to commonly utilized neutral atoms and trapped ions. To study this, we simulated the optical and mechanical modes of our device as the width ($w_1$ of Fig. 1a in the main text) was swept from from $\SI{600}{\nano\metre}$ to $\SI{2.3}{\micro\metre}$. We then calculated the photoelastic optomechanical coupling for each mechanical mode to the fundamental transverse electric optical mode as a function of device width. We present the result of this calculation for six mechanical breathing mode families (Fig. \ref{fig:bandwidth}b), where we have mapped the device-width parameter to the resonance frequency of the mechanical modes. Generally, as the device is made increasingly narrow, the resonance frequency for a given mechanical mode increases. However, for modes such as mode 3 that have strain almost entirely along the vertical axis, the resonance frequency hardly depends at all upon the device's width. We find that over frequencies ranging from $\SI{1}{\giga\hertz}$ to $\SI{5}{\giga\hertz}$, one can always find a mechanical mode exhibiting optomechanical coupling comparable to that of the device presented in the main text. In principle, one could achieve higher frequency operation by narrowing the device beyond $\SI{600}{\nano\metre}$, however difficulties with fabrication likely preclude this for now. At each simulated device width, we do not re-optimize the waveguide core width because this is computationally expensive and modifies the mechanical resonance frequencies. However, if one were targeting a specific frequency, the core width could also be varied to achieve even higher coupling.

\bibliographystyle{plain}  
\bibliography{sup_references.bib}